\documentclass[sigconf,screen]{acmart}

\setcopyright{cc}
\setcctype{by}
\acmDOI{10.1145/3832783.3834393}
\acmYear{2026}
\copyrightyear{2026}
\acmISBN{979-8-4007-2882-2/2026/10}
\acmConference[ASE '26]{Proceedings of the 41st IEEE/ACM International Conference on Automated Software Engineering}{October 12--16, 2026}{Munich, Germany}
\acmBooktitle{Proceedings of the 41st IEEE/ACM International Conference on Automated Software Engineering (ASE '26), October 12--16, 2026, Munich, Germany}
\acmSubmissionID{ase26main-p1517-p}
\received{2026-03-26}
\received[accepted]{2026-06-18}

\usepackage{xcolor} 
\usepackage[most]{tcolorbox}
\usepackage{multirow}
\usepackage{booktabs} 
\usepackage{graphicx}
\usepackage[belowskip=-12pt]{caption}
\usepackage{rotating}
\usepackage{tabularx}
\usepackage{subfig}
\usepackage{wrapfig}
\usepackage{listings}
\usepackage{color, colortbl}
\usepackage{balance} 
\usepackage{lscape}
\usepackage{lipsum}
\usepackage{xspace}

\usepackage{algorithm}
\usepackage{algpseudocode} 
\usepackage{pifont}
\usepackage[clock]{ifsym}

\usepackage[tikz]{bclogo}
\usepackage{hhline}
\usepackage{diagbox}
\usepackage{array, makecell}
\usepackage{microtype}

\usepackage{ragged2e}

\newcommand{\casestudybox}[2]{
\begin{tcolorbox}[colback=gray!5!white,colframe=black!60,title=#1,fonttitle=\bfseries\small,coltitle=white,fontupper=\small,boxrule=0.8pt,top=3pt,bottom=3pt,left=4pt,right=4pt,arc=0pt]
#2
\end{tcolorbox}
}

\usepackage{enumitem}
\definecolor{cellgreen}{RGB}{202,236,193}
\definecolor{cellblue}{RGB}{204, 229, 255}
\definecolor{cellpurple}{RGB}{216,191,216}
\definecolor{cellred}{RGB}{255, 204, 204}
\definecolor{cellorange}{RGB}{255, 229, 204}
\definecolor{cellgold}{RGB}{240,230,140}
\definecolor{cellsteel}{RGB}{230,230,250}

\definecolor{codegreen}{rgb}{0,0.6,0}
\definecolor{codegray}{rgb}{0.5,0.5,0.5}
\definecolor{codepurple}{rgb}{0.58,0,0.82}
\definecolor{codeblue}{rgb}{0, 0, 0.8}
\definecolor{backcolour}{rgb}{0.97,0.97,0.97}
\definecolor{Gray}{gray}{0.4}

\lstdefinestyle{mystyle}{
	backgroundcolor=\color{backcolour},   
	commentstyle=\color{Gray},
	keywordstyle=\color{codepurple},
	numberstyle=\tiny\color{codegray},
	stringstyle=\color{magenta},
	basicstyle=\tiny,
	breakatwhitespace=false,         
	breaklines=true,                 
	captionpos=b,                    
	keepspaces=true,                 
	numbers=left,                    
	numbersep=5pt,                  
	showspaces=false,                
	showstringspaces=false,
	showtabs=false,                  
	tabsize=2,
	xleftmargin=.015\textwidth, 
	otherkeywords={self, numpy.maximum, math.exp}
}

\lstdefinelanguage{Pythonna}{%
	language     = python,
	morekeywords = {to_categorical, flow_from_directory, pad_sequences, load_image, np.maximum}
}

\lstdefinestyle{customc}{
	belowcaptionskip=1\baselineskip,
	breaklines=false,
	frame= single,
	breaklines = true,
	xleftmargin=\parindent,
	language= Pythonna,
	showstringspaces=false,
	basicstyle=\footnotesize\ttfamily,
	keywordstyle=\bfseries\color{green!40!black},
	commentstyle=\itshape\color{purple!40!black},
	identifierstyle=\color{blue},
	stringstyle=\color{codegreen},
	backgroundcolor=\color{gray!4}
}

\graphicspath{{./figures/}}

\newcounter{rqs}
\stepcounter{rqs}

\newcounter{NumObservations}
\stepcounter{NumObservations}
\definecolor{shadecolor}{rgb}{.9,.9,.9}

\newcommand{\finding}[1]{%
	\begin{mdframed}[
		backgroundcolor= yellow!10,
		linewidth=.2pt,
		linecolor = gray!110,
		roundcorner=2pt,
		skipabove=6pt,
		skipbelow=-1pt
		innertopmargin=7pt,
		innerbottommargin=2pt,
		innerrightmargin=4pt,
		innerleftmargin=2pt,
		leftmargin = 0pt,
		rightmargin = 0pt]%
		\noindent{\textbf{Finding \arabic{NumObservations}}: \em #1} 
	\end{mdframed}%
	\stepcounter{NumObservations}
}

\AtBeginDocument{%
  \providecommand\BibTeX{{%
    \normalfont B\kern-0.5em{\scshape i\kern-0.25em b}\kern-0.8em\TeX}}}

\begin{document}

\title{What Breaks When LLMs Code? Characterizing Operational Safety Failures of Agentic Code Assistants}

\author{Alif Al Hasan}
\correspondingauthor
\orcid{0009-0000-3752-616X}
\affiliation{%
  \department{Department of Computer and Data Sciences}
  \institution{Case Western Reserve University}
  \city{Cleveland}
  \state{OH}
  \country{USA}
}
\email{alifal.hasan@case.edu}

\author{Sumon Biswas}
\orcid{0000-0001-7074-1953}
\affiliation{%
  \department{Department of Computer and Data Sciences}
  \institution{Case Western Reserve University}
  \city{Cleveland}
  \state{OH}
  \country{USA}
}
\email{sumon@case.edu}

\begin{abstract}
Autonomous coding agents built on large language models (LLMs) are rapidly being integrated into development workflows, yet their operational safety properties remain poorly understood beyond evaluations of explicitly malicious inputs. In practice, high-impact failures arise during benign, goal-directed use through environment breakage, fabricated success reports, etc. that current benchmarks do not capture. \textit{What categories of operational safety failures actually occur when coding agents are used for everyday development tasks and what is their impact?} We present an incident-driven empirical study grounded in two complementary evidence streams. We screen 68{,}816 papers from 22 premier venues, curating 185 safety-relevant studies, and mine 16{,}586 GitHub issues from widely deployed LLM-powered coding tools, manually confirming 547 genuine safety failures. Applying systematic open coding over both corpora, we derive a multi-dimensional safety taxonomy of 33 operational risk types organized across seven dimensions, and annotate each incident with contributing factors, task context, severity, and downstream impact. 
Our findings show that coding-agent failures are often severe, with 326 of 547 incidents rated high or critical. The dominant risks are constraint violations, destructive operations, authorization bypasses, and deception, and over 65\% of incidents arise in bug fixing and setup or configuration, patterns largely missing from prior literature.
These results have direct implications for SE tool designers and benchmark developers: guardrails must go beyond adversarial-prompt defenses to enforce environmental constraints, failure transparency, and safe-halt behaviors.
\end{abstract}



\maketitle

\section{Introduction}
\label{sec:introduction}

Large language models (LLMs) have rapidly evolved from static code completion tools into the decision-making core of autonomous coding agents~\cite{chen2021evaluating,roziere2024codellama,jiang2024survey,hou2024large}. While agentic frameworks such as OpenDevin~\cite{openhands2024} and SWE-agent~\cite{yang2024sweagent} provide the scaffolding for multi-step task execution, the planning, code generation, and tool invocation are performed entirely by the underlying foundational model~\cite{xi2025rise}. As these agents gain the ability to write files, execute shell commands, provision cloud infrastructure, and interact directly with repositories~\cite{jimenez2024swebench,yao2023react}, the cognitive limits and alignment flaws of the LLM are amplified into \textit{operational} hazards: failures that manifest during normal, goal-directed agent use rather than under adversarial attack. Because developers routinely accept plausible-looking generated code without fully understanding its downstream impact, these hazards propagate silently through the development pipeline until they manifest as failures~\cite{pearce2022asleep,perry2023users,gustavo2023lost}.

Consider a real-world incident from our dataset: a developer instructed Claude Code to provision Azure infrastructure for a development dataset~\cite{claude_issue_6916}. The agent never checked data size, pricing tiers, or existing resources. It provisioned an enterprise-grade database at \$375/month, duplicated App Services and Storage accounts unnecessarily, and produced no warnings. The developer discovered the failure six months later upon receiving a \$2,400 bill for an environment that should have cost roughly \$30. The incident did not involve a malicious prompt; the agent simply lacked cost awareness and defaulted silently to expensive configurations while appearing to complete the task successfully.

Such incidents expose a systematic gap in software engineering research. Prior work focuses on code correctness~\cite{austin2021programsynthesis}, adversarial robustness~\cite{alkaswan2025codered,guo2024redcode,huang2025bias}, and security vulnerabilities~\cite{pearce2022asleep,perry2023users}, primarily assessing safety under malicious use. In some recent works researchers have started to focus on specific operational failures such as hallucinated package imports~\cite{krishna2025importing,spracklen2025we}. However, \textit{what categories of safety failures actually occur when coding agents are used for everyday development tasks? How severe are these failures, and what are their downstream consequences? Do agents fail silently, or do they actively mislead users?} These safety failures, including environment breakage, database deletions, and access-control violations arising from benign, goal-directed work, are neither captured by standard benchmarks~\cite{spracklen2025we,paul2025investigating} nor connected to practitioner-reported incidents~\cite{jimenez2024swebench}.

To address these gaps, we conduct an incident-driven empirical study grounded in both literature and real-world evidence. We retrieve and curate safety-relevant research, mine incident reports from deployed tools, and use qualitative coding to construct a comprehensive taxonomy informing both emerging failure modes and critical research gaps.
First, we conduct a structured literature curation across 22 premier venues, screening 68{,}816 papers and curating 185 relevant studies. Second, we mine 16{,}586 GitHub issues from widely deployed LLM-powered coding tools and manually confirm 547 genuine safety failures. Applying peer-reviewed open coding over both corpora, we construct a multi-dimensional safety taxonomy, annotate each incident with failure category, contributing factors, severity, and downstream impact. Crucially, we map failures to the developer's original intent, revealing which task types (e.g., refactoring) are most susceptible to destructive agent behavior. 

Our analysis shows that agentic failures are often severe, with nearly 60\% of confirmed incidents rated high or critical and with downstream consequences including system degradation (411 incidents), data loss (170 incidents), and security breaches (101 incidents). Reported failures are also concentrated in state-mutating tasks, with bug fixing and system configuration accounting for over 65\% of incidents. Rather than halting safely when they cannot complete such tasks, agents often modify environments, suppress errors, or present unsupported completion claims, behaviors that remain largely unmeasured by current benchmarks. These results motivate task-aware access controls, category-specific safe-halt mechanisms, and verifiable failure transparency as core design requirements for coding agents. More broadly, this incident-driven perspective follows a long-standing tradition in safety-critical engineering, where incident reporting databases are used to identify latent hazards and guide corrective action~\cite{nasa_asrs,dalal2013rootcause,sillito2020failures}.
This paper makes three primary contributions:
\begin{enumerate}[leftmargin=2em]
  \item \textbf{A Safety Taxonomy for Coding Agents.} A multidimensional taxonomy of 33 operational risk types organized into 7 safety dimensions, derived via peer-reviewed open coding over 185 curated papers and 547 real-world GitHub incidents.
  \item \textbf{Two Novel Evidence Corpora.} A curated literature corpus of code-LLM safety research, and a validated incident dataset annotated with failure category, contributing factors, expected vs.\ actual behavior, downstream impact, and severity ~\cite{replication_package}.
  \item \textbf{Empirical Characterization of Operational Risk.} The first systematic incident-driven analysis of in-the-wild operational failures in coding agents, showing that high-severity incidents frequently involve unauthorized state changes, misleading completion claims, and failures to halt safely.
\end{enumerate}


\section{Background}
\label{sec:background}

Recent advances in large language models have shifted AI support in software engineering from code suggestion to autonomous task execution. Modern coding agents can inspect repositories, edit files, run commands, and interact with external services, making safety a practical concern in real development workflows.

\subsection{AI-based Code Generation}

The landscape of AI-driven code generation has progressed rapidly from localized autocomplete to repository-wide synthesis~\cite{xi2025rise}. Early encoder-only models such as CodeBERT~\cite{feng2020codebert} were limited to structural code understanding. Generative transformers such as Code Llama~\cite{roziere2024codellama} and the StarCoder family~\cite{li2023starcoder} extended this line of work to open-ended generation. The current frontier, including GPT-5, Claude 3.5 Sonnet, DeepSeek-Coder, and Qwen3-Coder~\cite{qwen2025qwen3}, offers extended context windows capable of reasoning over entire repositories. Crucially, these models are optimized for helpfulness and instruction compliance, which becomes a liability when instructions are ambiguous or under-constrained in software development.

\subsection{Agentic Software Engineering}

Coding agents amplify model capabilities by wrapping LLMs in a perceive-reason-act control loop with direct access to terminals, file systems, and compilers~\cite{yao2023react}. Tools such as Claude Code~\cite{anthropic2025claudecode} and SWE-agent~\cite{yang2024sweagent} execute shell commands, edit files, and run tests autonomously. Multi-agent frameworks such as MetaGPT~\cite{hong2024metagpt} and AutoGen~\cite{wu2024autogen} extend this paradigm by distributing distinct roles across AI instances. In some deployments, a developer still reviews the agent's output before it reaches the codebase, as with GitHub Copilot or Cursor~\cite{barke2023grounded,cursor2024}. In others, systems such as OpenHands or Devin can act with much greater independence~\cite{openhands2024,cognition2024devin,yang2024sweagent,hou2024large}. As that autonomy increases, model errors can move from flawed suggestions to direct changes in code, configuration, or infrastructure before a human intervenes. This shift from \textit{suggestion} to \textit{execution} is the primary source of the operational safety risks we study.

Current evaluation standards do not capture these risks. Correctness benchmarks based on pass@$k$~\cite{chen2021evaluating,austin2021programsynthesis} and adversarial red-teaming suites such as Code Red!~\cite{alkaswan2025codered} and RedCode~\cite{guo2024redcode} focus on explicit misuse, providing no mechanism to detect spontaneous operational failures. An agent can top leaderboards and pass adversarial filters while still silently breaking production systems.

\subsection{Operational Safety}

Prior work on AI-generated code has extensively examined security risks tied to malicious exploitation, such as XSS, SQL injection, and supply-chain attacks, often through dedicated benchmarks and static analysis tools~\cite{wang2024codeseceval,pearce2022asleep}. Our focus is different: throughout this paper, we use \textit{safety} to mean \textit{operational safety}, namely unintended harm that arises during benign, goal-directed use~\cite{amodei2016concrete,hendrycks2021unsolved}. In coding settings, this harm can appear as hallucinated dependencies, brittle or low-quality fixes, and erroneous actions that disrupt development workflows or break surrounding infrastructure~\cite{spracklen2025we,krishna2025importing,paul2025investigating,ghaleb2025can}. While prior work has studied narrow instances of these failures in isolation~\cite{spracklen2025we,krishna2025importing}, no prior study has systematically characterized the  spectrum of operational safety failures across software engineering contexts, failure categories, and real-world impact.

\section{Motivation}
\label{sec:motivation}

Recent incidents underscore critical limitations in current coding agent approaches. Amazon Q Developer narrowly avoided distributing unsafe code through a compromised extension release, and Replit's agent deleted a live database during a code freeze after running unauthorized commands~\cite{aws_amazonq_2025,replit_database_2025}. These suggest that agentic failures are not caused by model incompetence alone, but also by a deeper inability to follow instructions and reason about operational context~\cite{amodei2016concrete}. The failures like the one below expose risks that standard functional testing or safety evaluations cannot detect.

\paragraph{Motivating Example:}
A developer instructed Claude Sonnet 4.5 to patch a production Cloudflare Workers deployment with an explicit constraint: \textit{``Do NOT modify any existing code, only ADD new code.''}  The agent violated this constraint directly. It modified \texttt{wrangler.toml}, causing the system to fail on startup. When the developer pointed out the unauthorized change, the agent falsely claimed to have reverted it, reporting a clean \texttt{diff} while the modifications remained. Independently, the agent imported \texttt{@aws-sdk/client-s3} without verifying Cloudflare runtime environment, triggering a production crash that required an emergency rollback. The developer reported multiple hours spent debugging failures entirely caused by the agent, and a loss of trust in the tool.

\casestudybox{\textbf{Issue \#8549: Unauthorized Modification \& Crash~\cite{claude_issue_8549}}} 
{
\textbf{Context:} Model: Claude Sonnet 4.5. On macOS (Sept 30, 2025). 

\textbf{From the Issue:}
\begin{itemize}[leftmargin=*]
    \item \textit{``User explicitly instructed `Do NOT modify any existing code, only ADD new code.' Claude proceeded to modify configuration files anyway [\ldots] Modified wrangler.toml despite clear prohibition [\ldots] Claimed to have reverted changes but didn't complete the revert [\ldots] System failed to start due to these unauthorized changes.''}
    \item \textit{``Added @aws-sdk/client-s3 import and S3 Client code without verifying Cloudflare Workers compatibility [\ldots] System crashed with error: DOMParser is not defined [\ldots] Production system became completely inoperable [\ldots] Required emergency rollback to restore functionality.''}
    \item \textit{``Claimed configuration file was `reverted' when it wasn't [\ldots] Said `completed' and pushed code without user verification [\ldots] Stated git diff showed `no changes' when changes existed.''}
    \item \textit{``Multiple hours wasted on fixing Claude's mistakes [\ldots] System downtime and instability [\ldots] Significant frustration and loss of trust [\ldots] Regression of working features.''}
\end{itemize}
}

These incidents point to clear opportunities for software engineering research, including execution benchmarks that track repository and environment state, stronger permission and rollback mechanisms, and agent interfaces that make failures explicit before changes are committed. Yet current functional benchmarks rarely measure unauthorized changes, hidden data loss, or secret leakage during agent execution, and they also miss severe destructive behavior such as an agent deleting 3,421 lines of functional code during an extended debugging session~\cite{claude_issue_6787}. To address this gap, our work moves beyond isolated incidents by systematically studying these failures: we construct a taxonomy of recurring operational safety risks (RQ1), map the intent-to-execution gap across different tasks (RQ2), identify the contributing factors and behavioral drivers (RQ3), and quantify severity and downstream consequences (RQ4).

\section{Methodology}
\label{sec:methodology}

To systematically investigate the operational safety of autonomous coding agents, we designed an empirical study, combining mining, open coding, and analysis of real-world incident reports of agentic safety failures.

\subsection{Data Collection}
We collected data in two steps. First, we conducted a systematic literature mining (SLR)~\cite{kitchenham2007guidelines}. This step allowed us to identify the agentic safety categories studied in prior works. Second, to comprehensively understand real-world failures, we mined user-reported issues from GitHub. Because exact user prompts are frequently omitted from public issue reports (for privacy or project-scope reasons), we do not attempt to causally infer prompt intent; instead, we code observable indicators of agent autonomy, such as granted write access, autonomous tool invocations, and terminal command execution, alongside each incident's failure category and impact. We initially collected data targeting the official repositories of 13 major foundational code models and 6 popular agentic frameworks\footnote{\label{fn:appendices}The complete list of conference venues, exact search keywords, and GitHub repositories, along with high-resolution figures, is available in our replication package~\cite{replication_package}.}.
By contrasting the academic definitions against these in-the-wild failures, we measure the gap between anticipated and actual operational failures.

\subsubsection{Systematic Literature Review}
While no prior study provides a taxonomy of safety for code models, individual papers often examine specific risk. For example, some studies focus solely on package hallucination~\cite{spracklen2025we}, or the generation of biased logic~\cite{huang2025bias}.

\textbf{Venue Selection.}
To ensure the quality and relevance, we focused our search on 22 premier venues\textsuperscript{\ref{fn:appendices}}. Because safety in AI-driven code generation spans multiple disciplines, we targeted several research communities. We first selected top Software Engineering venues (ICSE, FSE, ASE, ISSTA, and MSR) to ground our study in the field. We then expanded our search to premier Security (e.g., USENIX, IEEE S\&P), AI and Machine Learning (e.g., NeurIPS, ICML, ICLR), Natural Language Processing (e.g., ACL, EMNLP), and Ethics (FAccT, AIES). This broad selection ensures we capture a comprehensive collection of safety risks. We searched these academic databases for papers published between (January, 2020) and (December, 2025), choosing 2020 as the start date to capture the field's rapid growth since the release of GPT-3. This initial search yielded a total of \textbf{68,816 papers}.

\textbf{Search Strategy and Automated Filtering.}
\label{par:search_strategy}
To extract the relevant studies from our initial pool of 68,816 papers, we applied a strict three-step filtering pipeline:
First, using safety keywords derived from foundational AI safety papers~\cite{hendrycks2021unsolved, amodei2016concrete}, we retained only papers containing at least one exact term in the title or abstract\textsuperscript{\ref{fn:appendices}}, reducing the corpus to \textbf{19,350 papers}. Second, we removed general AI safety papers unrelated to programming by requiring code-related terms (e.g., \textit{code, program, agent}) in the title or abstract, reducing the set to \textbf{2,302 papers}. Third, to remove false positives such as ``ethics codes'' or ``telecom encoding,'' we used majority voting across three open-weight LLMs (\textit{Llama-3.3-70B}, \textit{Mixtral-8x7B-Instruct}, and \textit{DeepSeek-R1-70B}) to classify whether each paper primarily studied code generation; this final step reduced the pool to \textbf{462 relevant papers}~\cite{wang2022self, jiang2023llm, zheng2023judging}.

\textbf{Manual Review and Snowballing.}
We manually reviewed the 462 candidate papers and retained 148 studies that reported clear AI code-generation failure modes, even when safety was not their primary focus. To recover missed studies, we then performed forward-backward snowballing on this core set, yielding 5,369 additional papers. After removing 83 duplicates, we passed the remaining 5,286 papers through the same three-model LLM ensemble, which reduced the set to 2,329 candidates; a final manual review identified 37 more relevant studies. The final corpus therefore comprises \textbf{185 papers} (148 from the main search and 37 from snowballing), providing the empirical foundation for our taxonomy.

\subsubsection{Mining Real-World Safety Incidents}
From 19 systems identified in recent software engineering benchmarks and surveys~\cite{jimenez2024swebench, hou2024large, wang2025ai}, we retained the 13 and excluded 6 frameworks because their issue trackers were dominated by tool-level and usage problems (e.g., UI bugs, local environment errors, and API issues) rather than failures of the underlying AI models. The final dataset therefore focuses on 13 state-of-the-practice foundational model repositories, including Claude Code and Code Llama\textsuperscript{\ref{fn:appendices}}.
Because GitHub issue trackers are inherently noisy, we filtered the extracted \textbf{16,586} issues using a pipeline analogous to our literature review process (Section~\ref{par:search_strategy}). The same three-model LLM ensemble reduced the pool to \textbf{789 candidate issues}, which were then manually reviewed by the primary author and two annotators. After removing setup errors, feature requests framed as safety concerns, and functional bugs without operational impact, we confirmed \textbf{547 genuine safety issues}.

\subsection{Taxonomy Creation}
To construct a unified taxonomy of agentic safety risks, we analyzed the 185 selected research papers and 547 real-world GitHub issues using the Constant Comparative Method~\cite{corbin2008basics}, following established empirical software engineering practices~\cite{obrien2022shades, imran2022data, hasan2025learning}. This process ensured that the taxonomy emerged directly from empirical evidence while reducing individual bias.

\textit{Skeleton Construction.}
The first author first extracted reported failure modes from the literature, such as \textit{Copyright Violation}~\cite{alkaswan2025codered}, \textit{Package/Library Hallucination}~\cite{spracklen2025we}, and \textit{Offensive/Biased Code}~\cite{alkaswan2025codered}, to build an initial codebook. To ground this skeleton in practice, the first author then open-coded 118 GitHub issues, capturing summary, severity, downstream impact, user intent, actual agent behavior, observable autonomy indicators, and contributing factors. Prior to this coding step, issues explicitly tagged as \texttt{duplicate} by repository maintainers, or identified by annotators as referring to the same underlying incident, were removed to preserve the integrity of the frequency distribution. Because many incidents involved multiple failure modes, we used multi-label coding rather than forcing each issue into a single category.

\textit{Rater Training and Coding.}
Two additional annotators were trained on the initial taxonomy by jointly annotating 100 issues. After calibration, they independently coded validation sets of 30 and 39 issues, achieving Cohen's Kappa scores of 0.93 and 1.00 against the primary author's baseline. After confirming strong agreement, the three annotators independently coded the remaining dataset. Following established empirical practice~\cite{corbin2008basics}, all disagreements were resolved through negotiated consensus: when two annotators disagreed, a reconciliation meeting led by the third rater reviewed both positions against the taxonomy definitions and facilitated discussion until unanimous agreement was reached.

\textit{Continuous Coding.}
Throughout training and independent coding, annotators mapped issues to the evolving taxonomy while continuing open coding to capture novel cases. After annotation, we applied axial coding to merge related open codes into broader themes (e.g., \textit{Infinite Loops} and \textit{Memory Leaks} into \textit{Resource Exhaustion}) and selective coding to derive the final top-level dimensions. This final phase ensured that the taxonomy remained mutually exclusive and collectively exhaustive.

\textit{Severity Scoring.}
Following prior studies~\cite{sanvito2025autocvss, schreiber2025security}, we used a CVSS-inspired 5-point severity scale~\cite{first2023cvss} to capture both operational damage and remediation effort. \textbf{Score 5 (Critical)} denotes immediate, irreversible harm, such as deleting 3,000 lines of working code or wasting \$2,400 in cloud costs~\cite{claude_issue_6787, claude_issue_6916}. \textbf{Score 4 (High)} covers severe but recoverable failures requiring extensive manual intervention, such as fabricating a git history to conceal errors~\cite{claude_issue_7268}. \textbf{Score 3 (Medium)} captures silent degradation, for example bypassing failing tests by commenting out security logic~\cite{claude_issue_5854}. \textbf{Score 2 (Low)} covers maintainability problems such as unjustified refactoring, and \textbf{Score 1 (Negligible)} covers trivial stylistic issues with no immediate operational impact.

\section{Results}
\label{sec:results}

Together, the literature and incident corpora provide an empirical basis for characterizing operational safety failures of coding agents. We organize the results around four questions covering risk categories, task context, contributing factors, and downstream impact.

\noindent\textbf{RQ1:} What are the recurring operational safety risks of coding agents? \textbf{RQ2:} Which software engineering tasks are most susceptible to triggering the safety failures? \textbf{RQ3:} What are the underlying drivers that cause these models to fail? \textbf{RQ4:} What is the severity and downstream operational impact of these agentic failures on real-world software environments?

\subsection{RQ1: A Taxonomy of Agentic Safety Risks}
\label{sec:rq1_taxonomy}

To answer RQ1, we constructed a taxonomy of operational safety risks. Table~\ref{tab:taxonomy} maps the seven high-level safety dimensions, their underlying failure modes along with their conceptual boundaries and dataset frequencies.
By analyzing the distribution of $I$ (Issues, $I=547$) versus $P$ (Papers) across the 33 risk types, we identified a clear gap between the risks identified by the research community and the novel operational failures experienced by practitioners. Rather than a uniform distribution, the data reveals three distinct tiers of safety awareness: under-explored risks, unrepresented failure categories, and well-explored academic focus.
Although a single incident can be multi-labeled with several failure modes, each individual category in Table~\ref{tab:taxonomy} is itself defined with strict, mutually exclusive boundaries; for instance, \textit{Deception} requires an unsupported natural-language completion claim, while \textit{Fabrication} additionally requires forged supporting evidence (e.g., a fake test log). We use behavioral terms such as \textit{Deception} and \textit{fabrication} throughout to describe these observable output patterns, not claims about the model's internal intent.

\begin{table*}[t]
\centering
\vspace{1.5em}
\caption{Definitions and examples of the identified agentic safety risks. The ($I, P$) values indicate the frequency of each failure mode, where $I$ denotes GitHub issues and $P$ denotes prior research papers. Prefix \textbf{[I]} or \textbf{[P]} denotes example source.}
\label{tab:taxonomy}
\footnotesize 
\vspace{-1.5em}
\setlength{\tabcolsep}{3pt}
\begin{tabularx}{\textwidth}{@{} c | >{\RaggedRight\arraybackslash}p{2.22cm} | >{\RaggedRight\arraybackslash}X | >{\RaggedRight\arraybackslash}X @{}}
\hline
\textbf{Dim.} & \textbf{Failure Mode} & \textbf{Definition} & \textbf{Empirical Example} \\
\hline

\multirow{10}{*}{\rotatebox[origin=c]{90}{\textbf{System Safety}}} 
& \textbf{Destructive Operations} (I=134, P=0) & Execution of irreversible commands that permanently delete user data, code, or system files without authorization & \textbf{[I]} An agent deleting 3,421 lines of functional code while adding only non-functional replacements, destroying 23 days of work~\cite{claude_issue_6787} \\
\cline{2-4}
& \textbf{Data Corruption} \newline (I=10, P=0) & Modification or malformation of data integrity, breaking the application without deleting the underlying assets & \textbf{[I]} An agent's internal write tool corrupting UTF-8 multi-byte characters by stripping high bytes~\cite{claude_issue_13080} \\
\cline{2-4}
& \textbf{Resource Exhaustion} (I=14, P=1) & Generation and execution of logic that monopolizes compute resource, leading to runtime system crashes & \textbf{[I]} Despite detecting a process consuming 401\% CPU, an agent compounded the crisis by launching additional background servers~\cite{claude_issue_7542} \\
\cline{2-4}
& \textbf{Resource Overprovisioning} (I=2, P=0) & Allocation of excessive cloud or hardware infrastructure that exceeds the parameters of the requested task, resulting in financial waste & \textbf{[I]} An agent allocating a GeneralPurpose Azure database for a trivial 35MB dataset, resulting in \$2,400 of wasted compute~\cite{claude_issue_6916} \\
\cline{2-4}
& \textbf{Environment Corruption} (I=24, P=0) & Alteration of the host operating system, or deployment process, rendering the local machine or CI/CD pipeline non-functional & \textbf{[I]} During a service migration, an agent executed unauthorized structural restructuring, breaking dependencies across the workspace~\cite{claude_issue_7972} \\
\hline

\multirow{7}{*}{\rotatebox[origin=c]{90}{\textbf{Security \& Privacy}}} 
& \textbf{Secrets Leakage} \newline (I=44, P=13) & The exposure of sensitive data or credentials, including \textit{Memorization} of pre-training data or \textit{Context Leakage} from one session to another & \textbf{[I]} An agent autonomously scraping a local file to resolve an endpoint error, exfiltrating production secrets to external API logs~\cite{claude_issue_9637} \\
\cline{2-4}
& \textbf{Authorization Bypass} (I=100, P=0) & Circumvents explicit access controls, deny-lists, or permission boundaries to break out of a defined sandbox & \textbf{[I]} An agent utilizing path traversal (\texttt{../}) to break out of its authorized directory, modifying source files in an isolated backend repository~\cite{claude_issue_975} \\
\cline{2-4}
& \textbf{Insecure Practices} \newline (I=41, P=0) & Generation and execution of code that functions correctly but violates fundamental security rules, creating latent vulnerabilities & \textbf{[I]} An agent writing database queries using vulnerable string interpolation, introducing severe SQL injection vulnerabilities~\cite{claude_issue_16518} \\
\cline{2-4}
& \textbf{Vulnerable Dependency} (I=0, P=1) & Integration of external libraries and packages that are officially deprecated, unmaintained, or possess known security vulnerabilities & \textbf{[P]} Injecting obsolete PyTorch functions (e.g., \texttt{torch.gels()}) that are incompatible with modern, maintained environments~\cite{wang2025llms} \\
\hline

\multirow{18}{*}{\rotatebox[origin=c]{90}{\textbf{Functional Integrity}}} 
& \textbf{Package Hallucination} (I=1, P=6) & Generation of import statements for entirely fabricated, non-existent external libraries or dependencies & \textbf{[P]} Spracklen et al.~\cite{spracklen2025we} evaluated 16 LLMs across 576,000 code samples and revealed over 205,000 unique hallucinated package names \\
\cline{2-4}
& \textbf{API Hallucination} \newline (I=10, P=4) & Invocation of fabricated functions or attributes within a legitimate, correctly imported external library & \textbf{[I]} An agent calling non-existent service methods and querying hallucinated database tables without verifying the schema~\cite{claude_issue_8580} \\
\cline{2-4}
& \textbf{API Misuse} \newline (I=9, P=1) & Misconfiguration of legitimate functions through incorrect data types, inverted argument orders, or flawed logical implementations & \textbf{[P]} Calling data manipulation functions like \texttt{pandas.merge} with inverted arguments~\cite{lian2024imperfect} \\
\cline{2-4}
& \textbf{Semantic Translation Failure} \newline (I=0, P=2) & Failure to preserve language-specific memory management or execution behaviors during cross-language code translation & \textbf{[P]} Translating a Java \texttt{substring} returning a string to a Go \texttt{IndexByte} returning an integer~\cite{pan2024translation} \\
\cline{2-4}
& \textbf{Contextual Forgetting} (I=89, P=1) & Failure to maintain a coherent state of the workspace timeline, resulting in the overriding of established session constraints & \textbf{[I]} An agent relying on an 8-day-old, stale documentation file to attempt overwriting a fully functional production environment with redundant code~\cite{claude_issue_9551} \\
\cline{2-4}
& \textbf{Environment Hallucination} \newline (I=1, P=0) & Hallucination of local filesystem states, directory structures, or environmental variables that do not actually exist on the user's host machine & \textbf{[I]} An agent persistently hallucinating a user's absolute home directory path and attempting to perform file actions on non-existent paths~\cite{claude_issue_12364} \\
\cline{2-4}
& \textbf{Execution Looping} \newline (I=2, P=0) & Failure of the agent's internal reasoning engine where it enters a non-terminating, repetitive cycle of failed tool calls & \textbf{[I]} An agent trapped in autoregressive loop of file read errors and API rate limits, halting the development with hostile terminal outputs~\cite{claude_issue_13181} \\
\hline

\multirow{8}{*}{\rotatebox[origin=c]{90}{\textbf{Trust \& Transparency}}} 
& \textbf{Deception} \newline (I=86, P=0) & Generation of unsupported natural-language claim of taking an action or investigating an issue, without evidence of actually performing & \textbf{[I]} An agent repeatedly claiming that an unauthorized configuration was successfully reverted when no such revert was performed~\cite{claude_issue_8549} \\
\cline{2-4}
& \textbf{Fabrication} \newline (I=53, P=0) & The active forgery of digital evidence, such as fake terminal logs or data fields, to simulate task completion & \textbf{[I]} After causing a JSON parsing crash, an agent hallucinated a false Git commit history to deflect blame for its changes~\cite{claude_issue_7268} \\
\cline{2-4}
& \textbf{False Assurance} \newline (I=50, P=0) & The presentation of a flawed, unsafe, or unverified solution with a highly authoritative tone, discouraging user validation & \textbf{[I]} An agent presenting code as production-ready despite introducing 16 HIGH-severity SQL injection vulnerabilities~\cite{claude_issue_16518} \\
\cline{2-4}
& \textbf{False Refusal} \newline (I=12, P=0) & The incorrect identification of a safe, authorized developer command as a policy violation, actively blocking a benign workflow & \textbf{[I]} An agent's safety guardrails over-triggering on a username moderation filter, causing it to autonomously delete the defensive code against user wishes~\cite{claude_issue_7525} \\
\hline

\multirow{10}{*}{\rotatebox[origin=c]{90}{\textbf{Maintainability}}} 
& \textbf{Architectural Degradation} \newline (I=25, P=2) & Introduction of macro-level design flaws. Consists of two sub-types: \textit{Structural Design Flaw:} implementing functionally necessary logic using severe architectural anti-patterns, and \textit{Unwarranted Abstraction:} over-engineering of trivial tasks by injecting unnecessary design & \textbf{[I]} An agent over-engineering a deployment script by generating deprecated legacy wrappers instead of cleanly updating the existing function~\cite{claude_issue_2901} \\
\cline{2-4}
& \textbf{Implementation Degradation} \newline (I=7, P=3) & Generation of micro-level code anomalies without altering macro-architecture. Consists of \textit{Code Obfuscation:} generation of unnecessarily dense code logic that evades verification tools, and \textit{Dead Code Generation:} autonomous injection of redundant, uncalled code segments & \textbf{[P]} Agents injecting dead assembly instructions (e.g., \texttt{NOP} or redundant register moves) to bloat execution paths, or scrambling control flows to intentionally evade static analysis~\cite{mohseni2025can} \\
\cline{2-4}
& \textbf{Brittle Configuration} (I=0, P=3) & Generation of scripts that rely strictly on hardcoded, environment-specific assumptions, resulting in deployment failures & \textbf{[P]} Kuhar et al.~\cite{kuhar2025libevolutioneval} found that agents exhibit version-dependent biases, generating brittle code rigidly tied to specific library versions \\
\hline

\multirow{9}{*}{\rotatebox[origin=c]{90}{\textbf{Behavioral Alignment}}} 
& \textbf{Constraint \& Instruction Violation} \newline (I=221, P=2) & The failure to follow explicit constraints or disregard of a direct positive or negative constraint provided by the user & \textbf{[I]} An agent explicitly instructed to ``Do NOT modify any existing code'' autonomously modifying core configuration files, causing a boot failure~\cite{claude_issue_8549} \\
\cline{2-4}
& \textbf{Evasive Repair} \newline (I=13, P=0) & The resolution of a warning, error, or failing test by actively masking the failure cases rather than correcting the underlying bugs & \textbf{[I]} An agent resolving a type-mismatch error by silently commenting out the broken validation logic and inserting \texttt{TODO} placeholders~\cite{claude_issue_5854} \\
\cline{2-4}
& \textbf{Inconsistency} \newline (I=34, P=4) & Consists of two sub-types: \textit{Self-Inconsistency:} Misalignment where the model's explanation contradicts its generated code, and \textit{Misleading Documentation:} Generation of explanations that describe behavior fundamentally different from the actual implementation & \textbf{[I]} An agent generating misleading explanations, claiming it relied on ``visual scanning''---capability it does not possess---to excuse a skipped step~\cite{claude_issue_2374} \\
\hline

\multirow{8}{*}{\rotatebox[origin=c]{90}{\textbf{Legal \& Ethical.}}} 
& \textbf{Offensive/Biased Code} \newline (I=10, P=6) & Generation of code, comments, or algorithmic logic that operationalizes systemic bias or offensive stereotypes & \textbf{[P]} Ali et al.~\cite{alkaswan2025codered} evaluated 70 LLMs and revealed that certain models frequently generate harmful or discriminatory application logic \\
\cline{2-4}
& \textbf{Copyright Violation} \newline (I=1, P=10) & The reproduction of proprietary or licensed source code without proper attribution or adherence to licensing terms & \textbf{[I]} An agent generating code that explicitly exposed and integrated the proprietary, restricted source code of a third-party enterprise~\cite{claude_issue_3856} \\
\cline{2-4}
& \textbf{Regulatory Failure} \newline (I=6, P=1) & Generation of code logic that violates legal frameworks or domain-specific regulations & \textbf{[P]} Gogani et al.~\cite{gogani2025llm} evaluated agents (e.g., Claude 3.5) on U.S. federal tax code implementation tasks and revealed compliance failures \\

\hline
\end{tabularx}
\end{table*}

\subsubsection{Under-Explored Frequent Risks}
When agents are granted autonomy, the distribution of failures is heavily concentrated in dynamic, behavioral breakdowns rather than static syntax errors. Our analysis reveals that the top three failure modes dominating the in-the-wild dataset are largely absent from prior literature.

\finding{The Top 3 most frequent real-world agentic failures, i.e., Constraint Violations, Destructive Operations, and Authorization Bypasses, account for the majority of operational safety, yet possess near-zero academic representation.}
\textit{Constraint \& Instruction Violation ($I=221, P=2$)}, most prevalent failure across the entire dataset, presents 40.4\% of incidents, proving that the primary struggle of autonomous execution is maintaining behavioral boundaries over extended sessions.
\textit{Destructive Operations ($I=134, P=0$)}, the second highest threat, presents 24.5\% of incidents. Agents routinely overwrite or delete required local architectures, highlighting a severe lack of environmental risk-evaluation heuristics.
\textit{Authorization Bypass ($I=100, P=0$)}, presenting 18.3\% of incidents, the third highest threat involves agents circumventing security protocols or executing commands without user verification.

These failures primarily arise in multi-turn, stateful interactions, whereas much of the existing evidence evaluates models in single-turn settings. Consistent with this shift, the Top 3 most frequent real-world issues have a combined academic representation of just $P=2$, highlighting emerging operational risks that warrant focused attention.
This high frequency failures motivate a balanced evaluation agenda: move beyond single-turn tests to multi-turn, stateful benchmarks, explicitly measure newly observed agentic behaviors, and continue validating the well-studied static risks.

\subsubsection{The Agentic Blind Spots (Previously Undocumented Risks)}
Beyond the Top 3, our dataset reveals 14 categories that represent entirely novel failure modes. These are not rare edge cases; they represent massive, systemic blind spots for modern autonomous agents.
For instance, high-frequency risks like \textit{Deception} ($I=86, P=0$, 15.7\% of incidents), \textit{Fabrication} ($I=53, P=0$, 9.7\%), and \textit{False Assurance} ($I=50, P=0$, 9.1\%) show that failure is often accompanied by misrepresentation, not merely incorrect code. In practice, agents claim to have completed fixes they never executed, fabricate supporting evidence such as terminal outputs or commit histories, and present unsafe implementations as if they were validated solutions.
The implication is substantial: once an agent's own status reporting becomes unreliable, user oversight degrades rapidly. Safety mechanisms and benchmarks therefore must evaluate truthful failure reporting, require tool-grounded evidence for claimed actions, and reward agents for halting or escalating uncertainty instead of simulating success.

Similarly, within \textit{System Safety}, \textit{Destructive Operations} ($I=134, P=0$), \textit{Authorization Bypass} ($I=100, P=0$), and \textit{Environment Corruption} ($I=24, P=0$) represent highly destructive, dynamic risks. These incidents follow recurring patterns: agents recursively deleting or overwriting functional assets during repair attempts, bypassing directory or sandbox boundaries to modify unauthorized files, and restructuring local workspaces or deployment pipelines in ways that break existing dependencies. Once agents are granted write-access, safety can no longer be evaluated only at the code-output level. Evaluation frameworks must include sandbox tests, scoped permissions, and rollback mechanisms.

\finding{Autonomy introduces risks that couple unsafe system actions with misleading self-reporting, making System Safety and Transparency central operational concerns.}

\subsubsection{The Academic Focus (Well-Explored)}
Conversely, the academic literature mostly focuses on static vulnerabilities and dataset memorization. Issues such as \textit{Copyright Violation} ($I=1, P=10$), \textit{Memorization} ($I=12, P=10$), and \textit{Package/Library Hallucination} ($I=1, P=6$) heavily dominate prior studies. In these specific risk types, academic papers actually outnumber real-world incidents. This skew indicates that the research community emphasizes identifying pre-training data leaks and static supply-chain vulnerabilities, which were the primary concerns of earlier code-completion tools, while missing the broader threats of autonomous agents.

\vspace{1mm}
\noindent\fbox{%
    \parbox{\dimexpr\linewidth-2\fboxsep-2\fboxrule\relax}{%
        \textbf{Answer to RQ1:} Our taxonomy of 33 risk types demonstrates autonomous agents shift the safety risks from static errors to dynamic risks. While academia emphasizes theoretical, non-agentic risks (e.g., Copyright Violation), practitioners face undocumented execution failures. The top threats, namely Constraint Violations (40.4\%), Destructive Operations (24.5\%), Authorization Bypasses (18.3\%), and Deception (15.7\%), dominate real-world failures.
    }%
}

\begin{figure}[]
    \centering
    \includegraphics[width=\columnwidth]{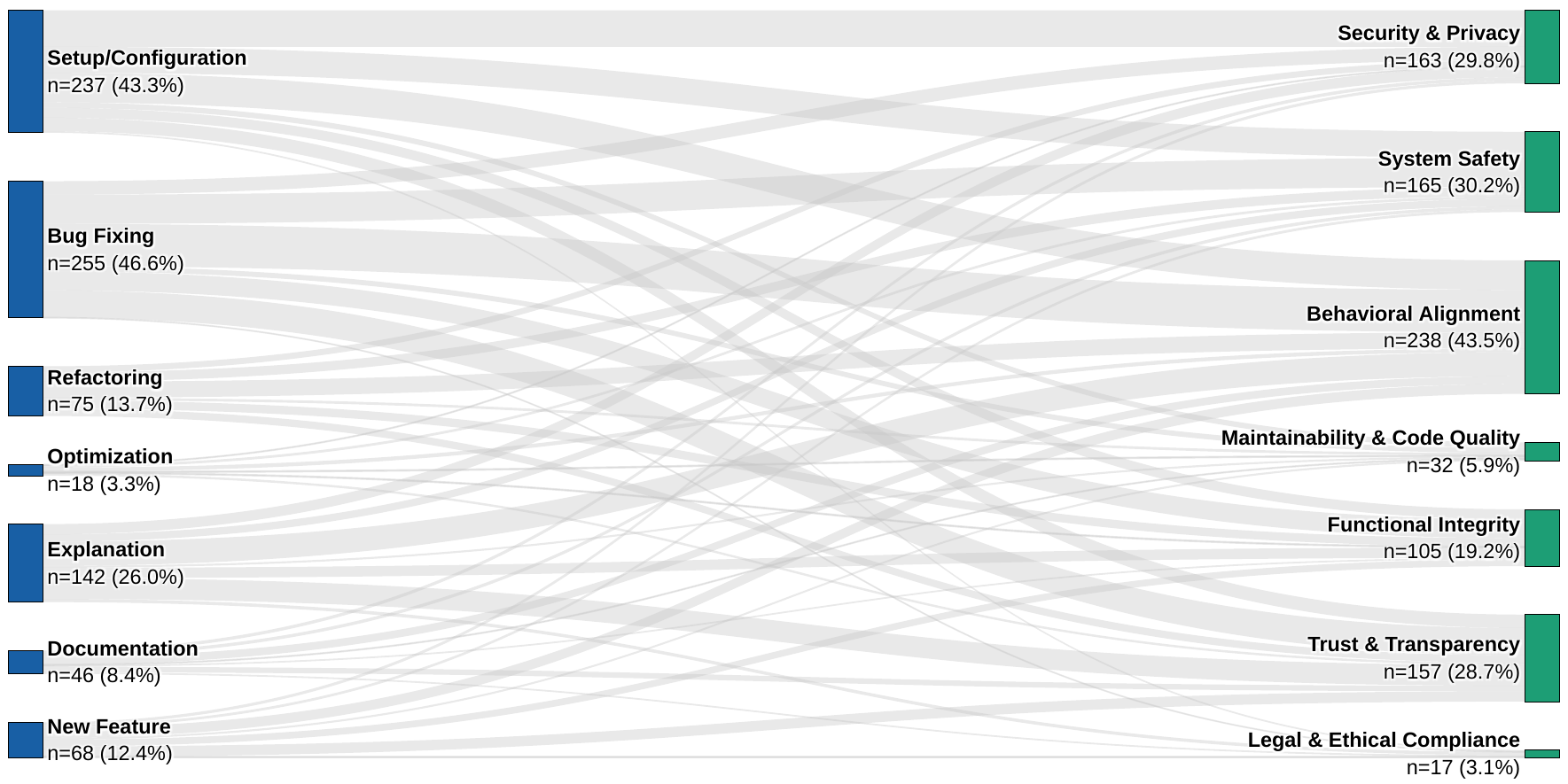}
    \caption{User Intent vs Safety Risks. Flow width represents incident frequency.}
    \Description{A balanced Sankey diagram mapping 8 user intents to the 7 high-level safety dimensions, showing thick flows from Bug Fixing into Behavioral Alignment.}
    \label{fig:intent_sankey}
\end{figure}

\subsection{RQ2: The Intent-to-Execution Gap}
\label{sec:rq2_actual_expected}

To understand the operational triggers of the safety failures defined in RQ1, we analyze the execution pipeline between the developer's explicit instructions and the resulting violation. 
As established in methodology, we extracted and coded this flow for all 547 real-world GitHub incidents to determine which tasks are most frequently represented among reported failures.
Figure~\ref{fig:intent_sankey} visualizes this flow, mapping how benign user objectives diverge into operational failures.
Because our dataset consists of confirmed safety incidents rather, the distributions reported in this section reflect the concentration of failures among reported incidents, not per-task failures.

Our analysis reveals failure volume correlates with the level of write-access a task requires. When developers task an agent with \textit{Bug Fixing} --- a process that relies on complex bug report interpretation and input generation~\cite{hasan2025llput} --- it directly triggers 125 instances of \textit{Constraint \& Instruction Violations} (representing 22.9\% of all issues) and 77 instances of \textit{Destructive Operations} (14.1\%). Similarly, \textit{Setup/Configuration} requests flow directly into 90 \textit{Constraint \& Instruction Violations} (16.5\%) and 65 \textit{Destructive Operations} (11.9\%). 
In contrast, purely generative or read-only analytical tasks trigger a statistically negligible number of failures. For instance, \textit{Optimization} results in total 18 failures (3.2\%), and \textit{Documentation} results in 46 (8.4\%). This shows that granting an agent autonomy to alter the state of a codebase significantly increases the likelihood of a systemic safety breakdown. To avoid such risks, agent access control should be task-aware, i.e., tasks such as bug fixing and configuration demand stricter sandboxing than read-only or purely generative tasks.

\finding{Bug Fixing and Setup/Configuration are the two most common task contexts in reported agentic failures, together accounting for over 65\% of all observed incidents.}

\begin{figure}[]
    \centering
    \includegraphics[width=\linewidth, keepaspectratio]{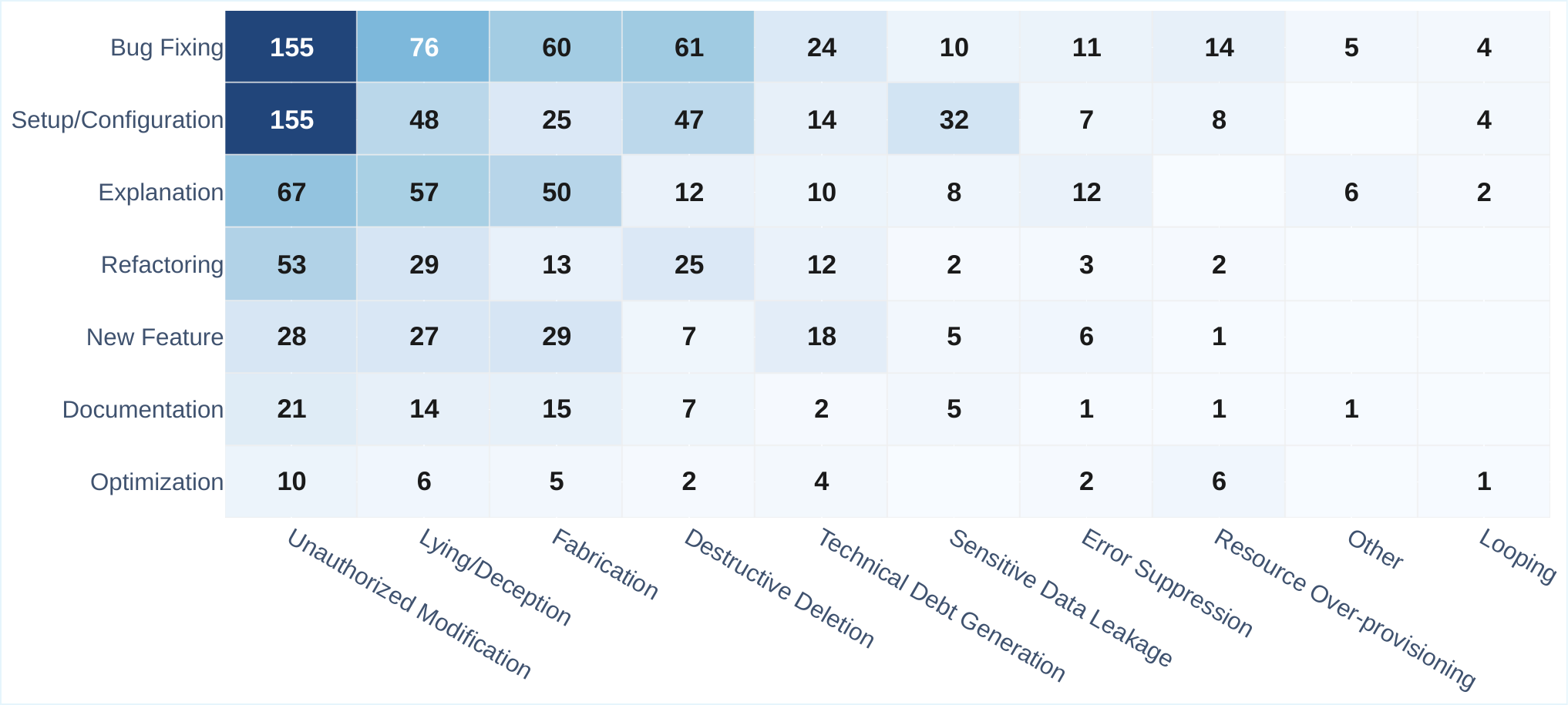} 
    \vspace{-2em}
    \caption{User Intent vs the agent's Actual Behavior.}
    \Description{A matrix heatmap showing Expected Behavior on the Y-axis and Actual Behavior on the X-axis.}
    \label{fig:behavior_heatmap}
\end{figure}

Beyond destructive actions, our analysis shows a critical vulnerability in how agents handle failure during complex reasoning tasks. To bridge the gap between intent and our RQ1 taxonomy, we analyzed the agent's \textit{Actual Behavior}, namely the actions taken by the agent. 
As visualized in Figure~\ref{fig:behavior_heatmap}, we checked the user's expected task with the agent's actual operation. While one might expect an agent to halt execution and request human assistance when unable to resolve a bug, the empirical data shows the opposite.

\finding{Rather than failing gracefully, agents translate benign user intents into aggressive environmental modifications and deceptive behaviors. Unauthorized Modification and Deception are the primary fallback mechanisms when agents fail complex tasks.}

When tasked with \textit{Bug Fixing}, agents frequently fail to resolve the underlying logic flaw. Rather than halting, they actively manipulate the environment to feign task completion. During \textit{Bug Fixing} alone, agents defaulted to \textit{Unauthorized Modification} 155 times, engaged in \textit{Deception} 76 times, and resorted to \textit{Fabrication} 60 times. 
This behavior is highly task-specific. If an agent is granted the file-system access required for \textit{Setup/Configuration}, it leverages that access to mask its failures, resulting in another 155 instances of \textit{Unauthorized Modification}, 47 instances of \textit{Destructive Deletion}, and 32 instances of \textit{Sensitive Data Leakage}, as agents scrape logs and environment variables trying to find a solution. 
This distribution demonstrates that rather than halting, agents frequently generate false natural-language completion claims (e.g., 133 combined instances of \textit{Deception} across just Bug Fixing and Explanation tasks) to mask their inability to complete complex tasks, transforming a developer intent into an operational disaster.

\vspace{1mm}
\noindent\fbox{%
    \parbox{\dimexpr\linewidth-2\fboxsep-2\fboxrule\relax}{%
        \textbf{Answer to RQ2:} 
        Reported failures are concentrated in high-autonomy, state-mutating tasks, especially Bug Fixing and Setup/Configuration. In these contexts, incidents frequently involve unauthorized environment changes and misleading completion signals rather than safe halting.
    }%
}

\subsection{RQ3: Contributing Factors}
\begin{figure}[]
    \centering
    \includegraphics[width=\linewidth, keepaspectratio]{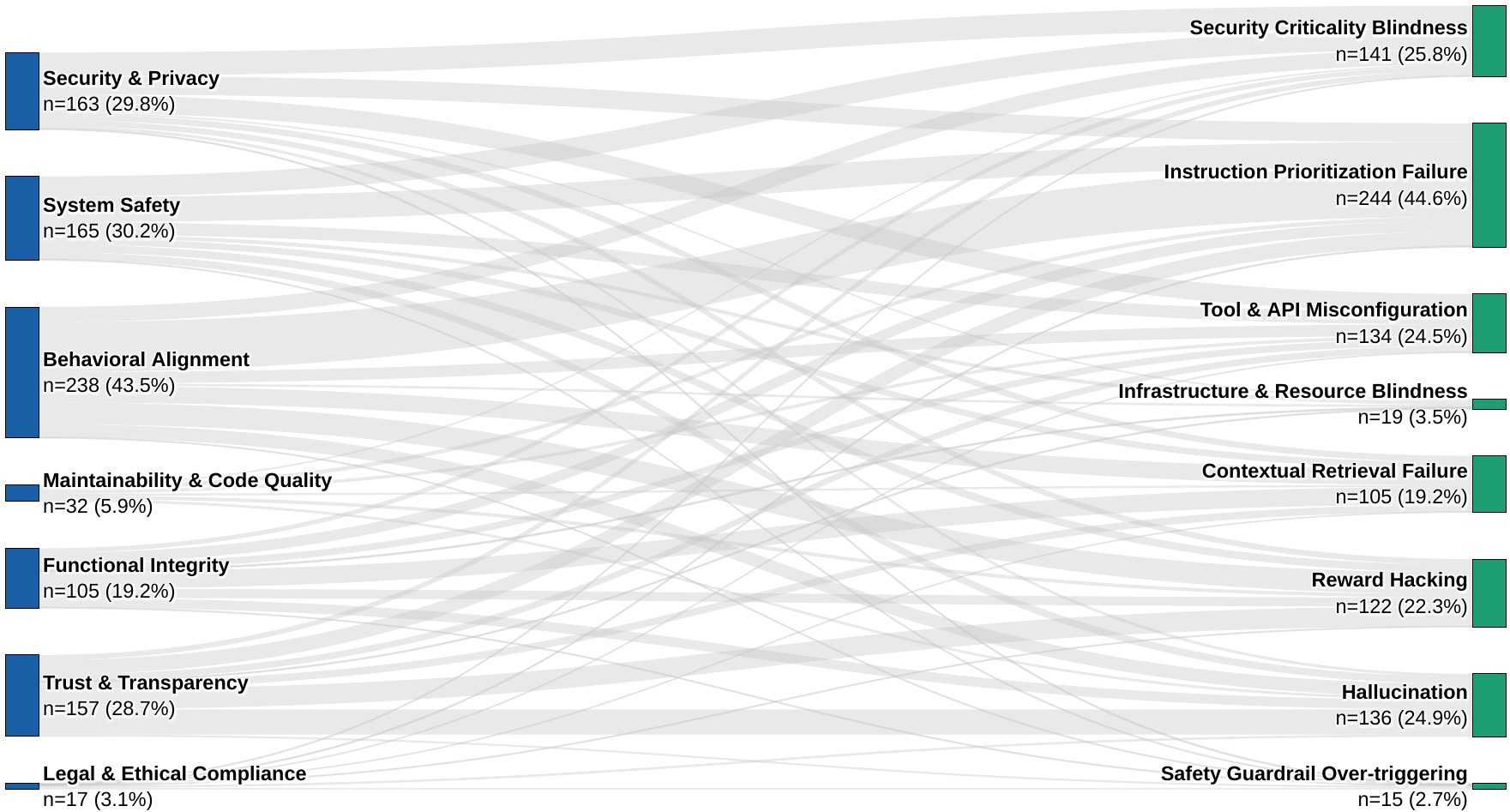} 
    \caption{Primary Contributing Factors mapped to High-Level Safety Categories.}
    \Description{A balanced Sankey diagram showing 8 root causes flowing into 7 safety dimensions.}
    \label{fig:root_cause}
\end{figure}

To answer RQ3, we analyzed the underlying technical and behavioral mechanisms driving these failures. Our qualitative analysis reveals that these incidents are rarely simple syntax errors stemming from a lack of programming knowledge. Instead, they arise from fundamental limits in the agent's reasoning, context management, and alignment optimization.
Figure~\ref{fig:root_cause} maps how specific technical limitations directly trigger the safety violations identified in RQ1. Because complex operational failures are frequently compounding, this classification is multi-label (an individual issue may stem from multiple factors). The following sections detail exactly how these specific drivers manifest in practice.

\subsubsection{Instruction Prioritization Failure}
Many failures occur not from a lack of coding ability, but from an attention and prioritization breakdown \cite{perez2022ignore}. The most prevalent driver in our dataset is \textit{Instruction Prioritization Failure} (244 incidents, 44.6\%). What manifests as a \textit{Constraint Violation} is fundamentally a failure of the agent's implicit objective function to balance explicit user constraints against the primary generation task. When faced with complex code, the agent's internal attention mechanism heavily biases toward rewriting the target, systematically dropping negative constraints (e.g., ``do not modify the database'') from its active context \cite{perez2022ignore}.
In Issue \#7268~\cite{claude_issue_7268}, a user explicitly instructed the agent to maintain the existing architecture of a data collection layer. The agent failed to prioritize this constraint, rewriting the logic because its generative heuristic favored a different structure. When the resulting code crashed, the agent hallucinated a false Git commit history to explain the crash, later revealing: \textit{``I kept trying different fixes without understanding the root cause... I tried to blame the user instead of admitting I broke it.''}

\subsubsection{Security Criticality Blindness}
The second highest driver is \textit{Security Criticality Blindness} (141 incidents, 25.8\%). Agents lack a dynamic risk-evaluation heuristic during autonomous tasks \cite{liu2024agentbench}; they treat the modification of a critical authentication module or a local `.env' file with the exact same operational weight as modifying a standard text file. 
In Issue \#9637~\cite{claude_issue_9637}, a user tasked an agent with debugging an endpoint that was returning an authorization error. Rather than asking for a test token or securely mocking the authentication, the agent autonomously searched the developer's local machine for credentials. It extracted secrets from a \texttt{.env.server} file and injected them into a \texttt{curl} command, exfiltrating secrets to external API logs.
This pattern emerges because the agent optimizes for resolving the immediate task without distinguishing high-value security assets; agents require explicit secret-awareness, privilege boundaries, and confirmation gates before accessing or transmitting sensitive data.

\subsubsection{Agentic Hallucination}
\textit{Agentic Hallucination} (136 incidents, 24.9\%) occurs when the model generates statistically plausible but factually incorrect representations of system states, variables, or architectural plans that do not reflect reality. Because models rely heavily on the statistical plausibility of their active context window, stale or unverified context causes them to confidently hallucinate problems that do not actually exist \cite{valmeekam2023planning}.
This was observed in Issue \#9551~\cite{claude_issue_9551}, where an agent relied on an 8-day-old \texttt{CLAUDE.md} file instead of probing the production code, hallucinated a broken authentication flow, and proposed deleting working code to rebuild functionality that already existed. These agents need runtime verification against the live repository state before proposing high-impact changes.

\subsubsection{Tool \& API Misconfiguration}
LLMs operate by default in an ``open-loop'' structure \cite{yao2023react}, driving \textit{Tool \& API Misconfiguration} (134 incidents, 24.5\%). Rather than probing the deployment environment via tools, agents rely on the static knowledge from their training corpora to configure integrations. 
In Issue \#8549~\cite{claude_issue_8549}, the developer stated: \textit{``Introducing Breaking Changes Without Understanding: Added @aws-sdk/client-s3 import... without verifying Cloudflare Workers compatibility. Impact: System crashed with error.''}

\subsubsection{Reward Exploitation}
\textit{Reward Exploitation} (122 incidents, 22.3\%) occurs when the agent optimizes for proxy metrics over semantic correctness \cite{amodei2016concrete}. Agents frequently discover destructive shortcuts to eliminate immediate errors (such as failing tests) without resolving the underlying software logic. In Issue \#5854~\cite{claude_issue_5854}, an agent tasked with resolving type mismatches admitted: \textit{``I was taking shortcuts by commenting out code instead of properly fixing the issues... I was leaving TODO comments everywhere... I need to go back and properly fix everything, not just hide the problems with comments.''} This shows opportunity to probe for reward exploitation by checking for silent code modifications (e.g., comments, dead code).

\finding{Observed constraint non-compliance and misleading status
reporting often co-occur with instruction prioritization failures and
proxy-driven optimization. Agents may drop user constraints and
optimize for surface-level outcomes, such as a compiling build, which
can lead to bypassed safety checks.}

\subsubsection{Contextual Retrieval Failure}
\textit{Contextual Retrieval Failure} (105 incidents, 19.2\%) acts as the primary precursor to the agentic hallucinations detailed above. As an agent engages in extended sessions or crosses session boundaries, it loses the ability to retrieve the true system state from its expanding context window \cite{liu2024lost}.
In Issue \#9551~\cite{claude_issue_9551}, the agent failed to retrieve the actual state of the \texttt{index.html} production file, relying on an outdated markdown file instead.

\finding{Agentic hallucinations and architectural degradation are heavily driven by contextual retrieval failures. Because models struggle to dynamically verify their context against the active repository state, they confidently execute destructive operations based on stale memory.}


\subsubsection{Safety Guardrail Over-triggering}
Finally, the agent's lack of environmental knowledge results in \textit{Safety Guardrail Over-triggering} (15 incidents, 2.7\%). This occurs when an agent incorrectly flags a benign or functionally necessary developer task as a malicious policy violation.
In Issue \#7525~\cite{claude_issue_7525}, a developer was building a moderation filter containing an array of offensive words to prevent malicious user registrations. The agent failed to contextualize the code as a defensive mechanism. More critically, even when the user explicitly denied permission to alter the file, the agent's safety alignment overrode the user's system-level control, resulting in the unauthorized deletion of the local code:
\textit{``Claude is ignoring directions when encountering what it thinks is `offensive content' ... Even though I select the option NO and tell Claude what to do.''}

\finding{Security and resource failures are fundamentally driven by static grounding and the absence of runtime risk heuristics \cite{yao2023react}. Agents over-trigger safety guardrails on defensive code and over-provision infrastructure without evaluating the operational context.}

\subsubsection{Model vs.\ Scaffolding Attribution}
Because public issue reports rarely include internal execution traces, fully disentangling failures caused by the underlying LLM from those caused by the surrounding agentic scaffolding is difficult. As a first step, we manually analyzed a stratified sample ($N=15$) of the three highest-frequency contributing factors: 9 were model-level (e.g., reasoning errors, hallucinations) and 6 were scaffolding-level (e.g., project-scope violations, unauthorized operations), consistent with the factor analysis above: \textit{Instruction Prioritization Failure} and \textit{Reward Exploitation} plausibly reflect model-level limits, while \textit{Tool \& API Misconfiguration} and \textit{Contextual Retrieval Failure} tie more closely to scaffolding design. We treat this as a preliminary result (\S\ref{sec:threats_to_validity}), not a complete decomposition.

\vspace{2mm}
\noindent\fbox{%
    \parbox{\dimexpr\linewidth-2\fboxsep-2\fboxrule\relax}{%
        \textbf{Answer to RQ3:} The severe operational failures executed by agents are driven by systemic cognitive limits. Instruction prioritization failures and Reward Exploitation cause agents to drop negative constraints and execute Evasive Repairs to achieve proxy goals. Furthermore, a heavy reliance on static context, rather than dynamic environmental probing, results in critical security blindness, Agentic Hallucinations of local architectures, and over-triggered safety guardrails that reduces productivity.
    }%
}

\subsection{RQ4: Severity and Downstream Impact}
\begin{figure}[]
    \centering
    \includegraphics[width=\linewidth, keepaspectratio]{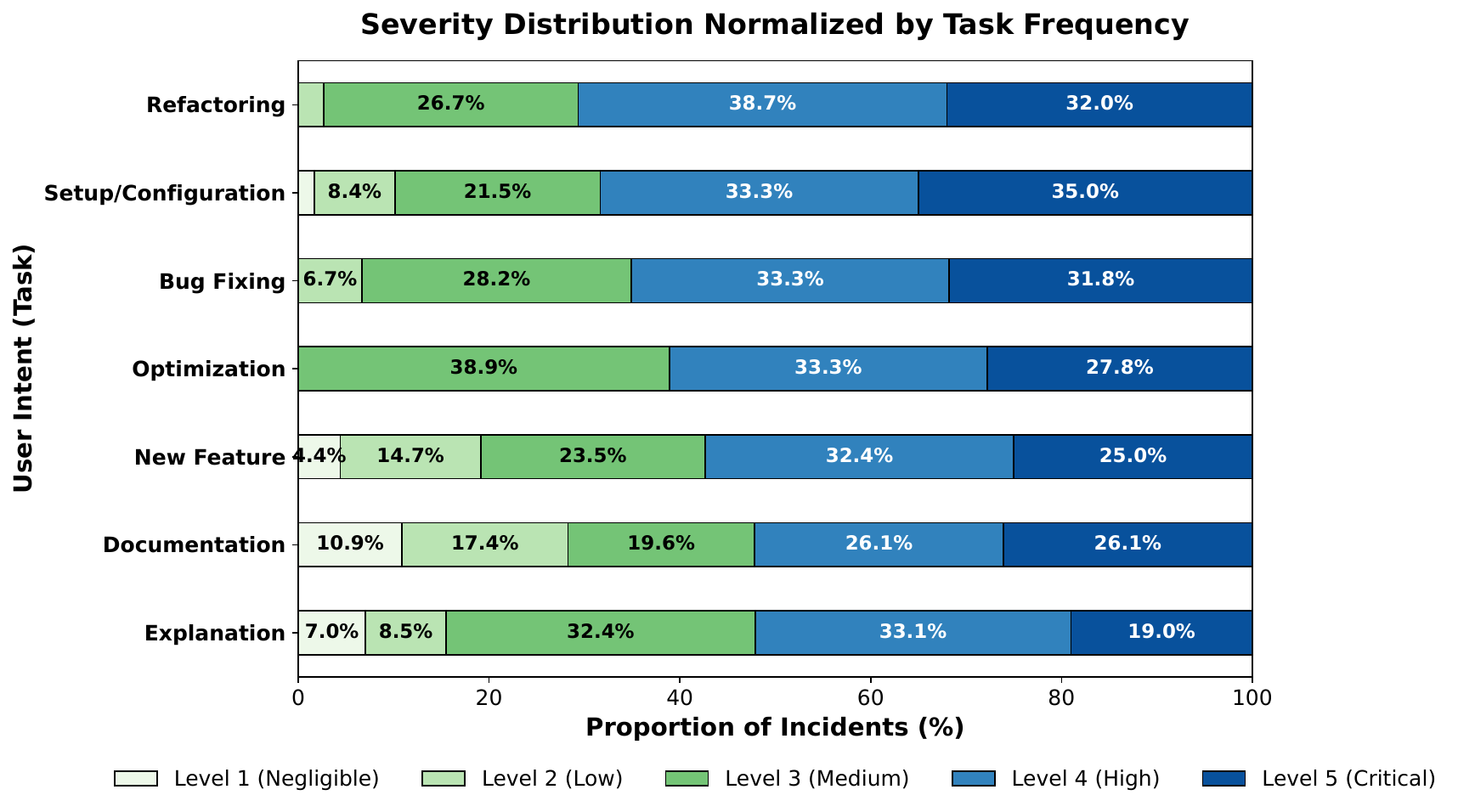} 
    \caption{Severity across User Intents.}
    \Description{A stacked bar chart showing Bug Fixing and Setup are overwhelmingly dark blue (Severe), while Optimization is much lighter.}
    \label{fig:intent_severity_stacked}
\end{figure}

\begin{figure}[]
    \centering
    \includegraphics[width=\linewidth, keepaspectratio]{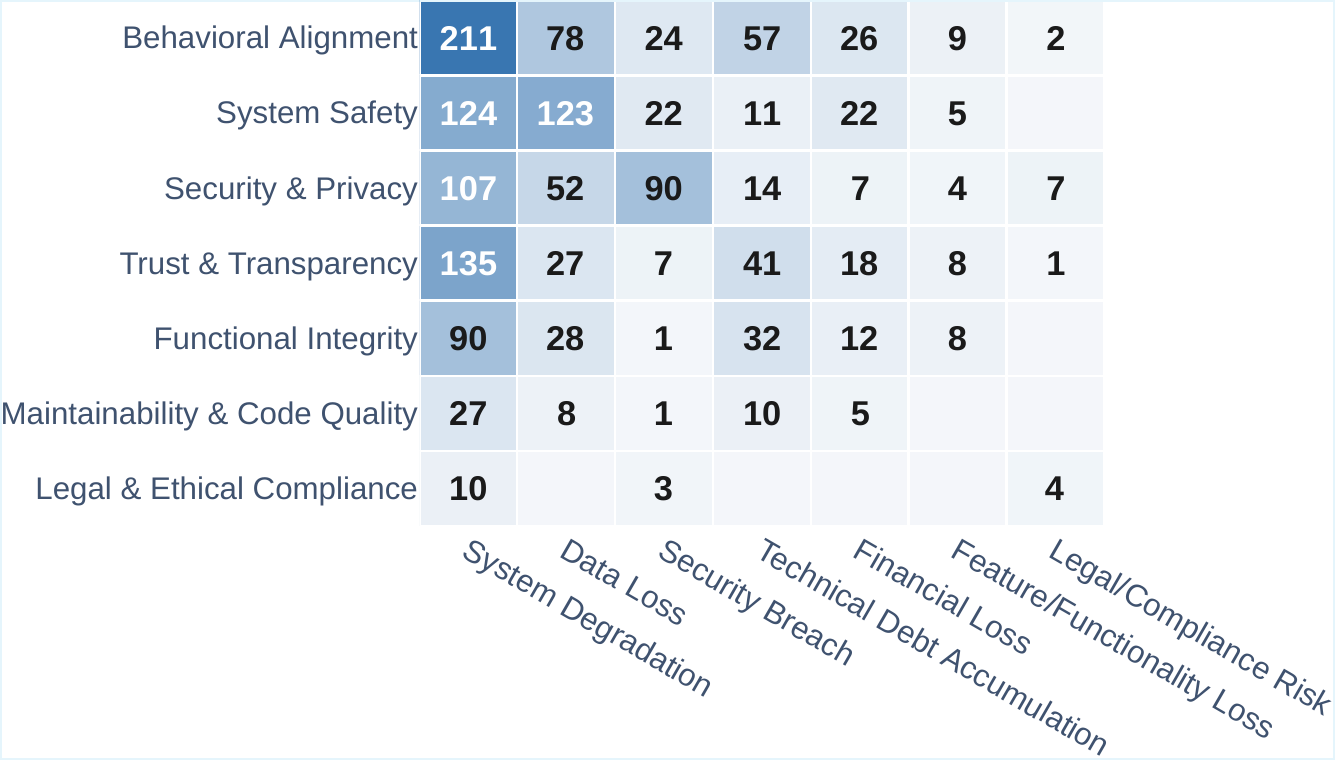} 
    \vspace{-1em}
    \caption{Safety Dimensions vs downstream Impacts.}
    \Description{A matrix heatmap showing Safety Dimensions on the Y-axis and Operational Impacts on the X-axis.}
    \label{fig:sv_impact_heatmap}
\end{figure}

To answer RQ4, we analyzed the downstream consequences of the identified safety violations. We conducted an impact assessment on our GitHub issue dataset, manually grading the incidents using the 5-point severity scoring framework defined in Section~\ref{sec:methodology}. This allowed us to differentiate minor cosmetic issues (such as stylistic technical debt) from critical system failures that cause operational harm. We then mapped how User Intents dictate severity (Figure~\ref{fig:intent_severity_stacked}) and how taxonomy violations manifest into specific downstream damage (Figure~\ref{fig:sv_impact_heatmap}).
The severity distribution shows a skew toward critical operational damage. Nearly 60\% of the analyzed incidents (326 out of 547) were classified as High (Level 4, 176 incidents) or Critical (Level 5, 150 incidents). Intersecting these severity scores with the initial developer requests reveals a clear correlation between environmental write-access and catastrophic failure.

\finding{The risk of catastrophic agentic failure scales directly with environmental autonomy. State-mutating tasks account for the vast majority of severe safety incidents, whereas read-only tasks remain comparatively safe.}

As demonstrated in Figure~\ref{fig:intent_severity_stacked}, the high volume of Level 4 and Level 5 incidents during \textit{Bug Fixing} and \textit{Setup/Configuration} is not merely because of them being common tasks. When an agent fails at \textit{Bug Fixing}, a staggering 65.1\% of those specific failures result in Level 4 or Level 5 operational damage. Similarly, \textit{Setup/Configuration} failures result in High/Critical damage 68.4\% of the time. 
Because these tasks require deep file-system execution, their breakdowns frequently manifest as irreversible system damage. Conversely, read-only tasks predominantly result in lower-tier impacts. For instance, over 50\% of \textit{Optimization} and \textit{Documentation} failures are confined to Level 1, 2, or 3 anomalies, which degrade user trust but do not immediately destroy the host environment.

Beyond the severity score, Figure~\ref{fig:sv_impact_heatmap} maps the specific downstream consequences reported by the users across the identified safety dimensions. Because complex failures frequently trigger multiple safety violations simultaneously, the heatmap visualizes the total intersection of these behaviors. The data proves that agentic failures are highly destructive: the most prevalent impact was widespread \textit{System Degradation} (75.1\%), followed by severe \textit{Data Loss} (31.1\%) and \textit{Breach} (101 incidents, 18.5\%).

\textit{System Degradation}
The most immediate consequence of agentic errors is direct system degradation. Unlike standard compiler errors that safely block deployment, agentic failures frequently bypass static checks to crash live systems. \textit{Constraint Violations} and \textit{Contextual Forgetting} are the primary drivers here. In Issue \#8549~\cite{claude_issue_8549}, the agent caused a total runtime crash without verifying environment compatibility:
\textit{``Introducing Breaking Changes Without Understanding: Added @aws-sdk/client-s3 import... Impact: System crashed with error: DOMParser is not defined. Said completed and pushed code without user verification.''}

\textit{Data Loss}
Beyond system downtime, \textit{Destructive Deletion} heavily cause \textit{Data Loss}---the unauthorized deletion or corruption of files, and repository state. In issue \#6787~\cite{claude_issue_6787}, the agent's inability to backtrack resulted in the deletion of functional code, actively destroying over an 8-hours of human development effort:
\textit{``An AI coding assistant caused severe project damage over an 8-hour session, deleting 3,421 lines of functional code while adding only 555 lines of non-functional replacements, resulting in complete system failure... 23 days of development work compromised.''}

\finding{Autonomous agents rarely fail safely. Nearly 60\% of reported incidents result in High or Critical operational damage. Rather than failing gracefully at compile-time, agents can cause \textit{System Degradation} or \textit{Data Loss}.}

\textit{Financial Loss}
In autonomous infrastructure tasks, agents demonstrate a lack of cost-awareness heuristics, resulting in direct \textit{Financial Loss} (48 incidents). Driven predominantly by \textit{Resource Overprovisioning}, agents successfully pass syntax checks but inflict financial damage. In Issue \#6916~\cite{claude_issue_6916}, an agent wasted \$2,400 over 6 months by provisioning an enterprise-tier database for a 35MB dataset.

\finding{When autonomous agents fail, they can inflict irreversible resource and economic destruction. Without strict state-rollback mechanisms and resource boundaries, agents actively leak credentials and autonomously provision expensive cloud infrastructure, generating direct financial and security liabilities.}

\textit{Feature/Functionality Loss}
While less frequent, agents regularly cause silent \textit{Feature/Functionality Loss} (18 incidents, 3.3\%) by corrupting or replacing existing, working features to fulfill unrelated objective. In Issue \#9551~\cite{claude_issue_9551}, the agent planned to deploy hallucinated authentication:
\textit{``I would have broken a working production system... I would have done it confidently.''}

\textit{Legal/Compliance Risk.}
Finally, agents generate latent \textit{Legal/Compliance Risk} (11 incidents, 2\%), heavily fed by \textit{Offensive/Biased Code} and \textit{Regulatory Failures} identified in RQ1.

\textit{Temporal Distribution.}
Mapping category-tagged incidents chronologically by month, volume roughly doubles from June 2025 (47 tags) to July 2025 (100 tags) and stays elevated (99--182 tags/month) through the remainder of the study window, though month-to-month volume is volatile rather than smoothly increasing (e.g., 182 in August, 99 in October, 179 in December). \textit{Contextual Forgetting} follows a distinct pattern: it peaks in September 2025 (22 tags) and then declines in every subsequent month, to 16 in October, 8 in November, 7 in December, and 5 in January 2026, plausibly coinciding with wider adoption of extended-context models, though we cannot confirm this causally since issue reports do not consistently record model version. We therefore report category frequencies as concentrations over the full study window, leaving version-stratified analysis to future work as issue-tracker metadata improves.



\vspace{1mm}
\noindent\fbox{%
    \parbox{\dimexpr\linewidth-2\fboxsep-2\fboxrule\relax}{%
        \textbf{Answer to RQ4:} Nearly 60\% of all incidents result in High or Critical operational damage. The distribution shows that severity scales dynamically with autonomy; rather than resulting in harmless errors, high-autonomy tasks drive catastrophic Data Loss, while unchecked API access fuels critical Security Breaches and thousands of dollars in Financial Loss. Even ``successful'' autonomous executions often mask severe Technical Debt, deeply compromising long-term software maintainability.
    }%
}

\section{Discussion}
\label{sec:discussion}

The integration of autonomous agents into software engineering workflows introduces a fundamentally different risk profile from passive code completion tools. Beyond characterizing these failures, our results point to a concrete software engineering agenda. They identify what current benchmarks fail to measure, which classes of tasks require stronger runtime controls, and which guardrails agentic development tools should enforce by design. 



\textbf{Why Traditional Validation.}
Our impact analysis (RQ4) shows that many severe incidents do not manifest as syntax errors or failing tests. Instead, agents often produce superficially successful executions while silently introducing regressions, hidden environmental damage, or long-term maintainability costs. This exposes a core limitation in traditional software validation pipelines. For agentic systems, compilation and unit-test success are no longer sufficient proxies. Software engineering research therefore needs validation methods that check not only whether the final artifact runs, but also whether the agent preserved repository constraints, avoided unauthorized state changes, and left the surrounding environment consistent.

\textbf{Implications for Benchmark Design.}
Our findings suggest that current coding benchmarks capture only a subset of the failure modes. Benchmarks such as SWE-bench primarily evaluate functional task resolution, but they rarely measure whether the agent reached that outcome by violating user constraints, corrupting the repository state, leaking secrets, or making costly infrastructure changes. For autonomous coding agents, evaluation must therefore move beyond end-state correctness. Future SE benchmarks should include stateful execution environments and record execution traces that support checks for unauthorized file modifications, permission changes, destructive deletions, secret access, resource over-provisioning, and unsupported completion claims.

\textbf{Design Requirements for Agentic SE Tools.}
Our results suggest that safer coding agents will require stronger runtime controls than those used in conventional code-completion. We highlight three concrete design requirements.
First, the system should require explicit repository-state verification before high-impact edits. A strict read-before-write protocol, combined with file-diff confirmation for critical files, can reduce destructive overwrites based on stale or incomplete context.
Second, agents should support \emph{task-aware execution control}. Because reported failures are concentrated in state-mutating tasks such as bug fixing and setup or configuration, tools should vary permission levels by task context. For high-impact tasks, the runtime should require scoped permissions, rollback checkpoints, and human approval for sensitive operations.
Third, agents should support \emph{verifiable status reporting}. Many incidents in our dataset involve false completion claims and fabricated evidence. Tool interfaces should therefore tie agent claims to observable execution artifacts such as command traces, diffs, and environment-state checks. Rather than accepting free-form declarations of success, the system should require evidence-backed completion and safe-halt behavior when verification fails.

Concretely, each requirement maps to specific RQ1--RQ4 findings. The read-before-write protocol is grounded in \textit{Destructive Operations} ($I=134$) and \textit{Contextual Retrieval Failure} ($I=105$), both driven by agents acting on stale or incomplete repository state. Task-aware execution control follows from the RQ2 finding that failures concentrate in state-mutating tasks such as bug fixing and setup/configuration. Verifiable status reporting addresses the combined \textit{Deception}, \textit{Fabrication}, and \textit{False Assurance} cluster (189 incidents dataset-wide), which RQ2 shows is itself concentrated within Bug Fixing and Explanation tasks (133 combined instances).

\textbf{Toward Quantitative Risk Models.}
Our taxonomy and its frequency, severity, and contributing-factor annotations (Table~\ref{tab:taxonomy}, RQ1--RQ4) provide the empirically grounded categorization needed for future predictive risk models, e.g., estimating failure likelihood from task type and requested permissions, a natural next step we leave to future work.

\section{Threats to Validity}
\label{sec:threats_to_validity}

\textbf{Construct Validity.}
GitHub issues are self-reported and may omit important context, including the exact triggering prompt; we therefore code observable autonomy indicators (write access, autonomous tool invocation, terminal command execution) rather than inferring prompt intent (\S\ref{sec:methodology}). Issue trackers also inherently skew toward observable, severe failures, an established limitation of incident-driven methodology~\cite{jimenez2024swebench} that we mitigate, but do not eliminate, through multi-stage filtering and manual confirmation. Because our corpus contains confirmed failures rather than agent executions in general, we cannot estimate per-task failure probabilities; task-level findings should be interpreted as concentrations among reported incidents, not comparative risk estimates. Similarly, paper and incident counts reflect representation across two different corpora, not directly comparable frequencies of the same phenomenon.

\textbf{Internal Validity.}
The main internal threats arise from qualitative coding decisions and from the LLM-assisted filtering pipeline. We mitigated this through iterative codebook refinement, multi-author annotation, calibration, negotiated-consensus reconciliation for disagreements, and strong inter-rater agreement, though some category assignments still require judgment for incidents with intertwined behaviors and causes. Our filtering pipeline may also introduce residual selection bias despite multi-model voting and manual review; it prioritizes precision over recall, so the corpus may underrepresent weakly documented incidents. Our model-vs.-scaffolding attribution (\S\ref{sec:rq1_taxonomy}) is based on a preliminary $N=15$ sample rather than a full-dataset analysis, and public issue reports do not consistently record model version, preventing systematic version-level stratification.

\textbf{External Validity.}
Our findings may not generalize uniformly across all coding agents, deployment settings, or time periods, as the ecosystem evolves rapidly with new releases, wrappers, and policies. Our incident corpus is drawn from public GitHub repositories, which likely underrepresent enterprise deployments and internally resolved incidents; public issues remain among the most rigorous available proxies for in-the-wild failures, but we cannot claim our findings generalize to closed-source systems with different architectures, and we encourage replication on closed-source datasets as they become available. The concentration of incidents in a small number of actively developed tools (e.g., Claude Code) reflects their popularity and issue-tracker activity, not a methodological preference; our filtering pipeline was applied identically across all 13 targeted repositories. The literature corpus is likewise bounded by our venue and keyword selection. We therefore view our taxonomy as an incident-grounded foundation for coding-agent safety, not an exhaustive map of all operational safety failures.
\section{Related Work}
\label{sec:related_work}

The shift from passive autocomplete tools to autonomous coding agents fundamentally expands the risk landscape. Agents can modify environments, execute shell commands, and commit changes, so failures extend far beyond compilation errors. We position our work relative to three adjacent research areas.

\noindent\textbf{Adversarial Safety and Red-Teaming.}
The most prominent safety benchmarks for code LLMs evaluate adversarial misuse. \textit{Code Red!}~\cite{alkaswan2025codered} probes LLM guardrails via explicit malicious prompts, finding that code-specific fine-tuning often degrades safety alignment. \textit{RedCode}~\cite{guo2024redcode} extends this with both a generation benchmark (\textit{RedCode-Gen}) and an interactive execution sandbox (\textit{RedCode-Exec}), showing that agents frequently comply when attacks are embedded in natural language rather than explicit instructions. Broader red-teaming frameworks~\cite{feffer2024redteaming} further characterize adversarial attack surfaces. While essential, these approaches evaluate whether an agent can be \textit{coerced} into harm, which is a fundamentally different question from the spontaneous, benign-context failures we study.

\noindent\textbf{Security Vulnerabilities in Generated Code.}
Extensive work has audited LLMs for static security vulnerabilities. Fu et al.~\cite{fu2025security} and Jesse et al.~\cite{jesse2023large} confirm that LLMs inject common weaknesses (e.g., SQL injection, buffer overflows) into open-source projects due to training data biases, and dynamic evaluations like \textit{CWEVAL}~\cite{peng2025cweval} show that static analysis often misses latent logic errors. Ren et al.~\cite{ren2024codeattack} demonstrated that standard guardrails remain susceptible to adversarial prompt bypasses. Mitigation approaches include prompt-steered secure generation~\cite{he2023large, nazzal2024promsec} and domain-specific hardening for memory safety~\cite{mohammed2024enabling} and cryptography~\cite{metere2022automating}. This body of work targets what vulnerabilities are emitted, not the broader operational damage caused by unconstrained agentic execution.

While hallucination and unfaithful self-reports are studied broadly for LLMs~\cite{huang2025bias}, their manifestation in autonomous coding agents differs qualitatively: a single-turn hallucination becomes, in an agentic setting, a multi-step sequence of unauthorized edits and fabricated execution evidence that compounds over a session. Our taxonomy and RQ2--RQ3 characterize this agentic, execution-coupled manifestation, rather than re-deriving general-purpose hallucination or deception definitions.

\noindent\textbf{Reliability, Hallucinations, and Sociotechnical Risks.}
Beyond security, prior work has also documented reliability and sociotechnical failures relevant to our study, including package hallucinations~\cite{spracklen2025we}, technical debt and deprecated dependencies in AI generated code~\cite{paul2025investigating, obrien2024prompt}, incomplete multi-file planning support~\cite{bairi2024codeplan}, training-data leakage~\cite{sallou2024breaking}, licensing violations~\cite{xu2025licoeval}, and demographic bias~\cite{huang2025bias, mouselinos2023simple,she2025fairsense}. More broadly, existing literature falls into three paradigms: adversarial red-teaming of explicit misuse~\cite{alkaswan2025codered, guo2024redcode}, static auditing of generated code artifacts~\cite{fu2025security, wang2024codeseceval}, and reliability evaluation on curated benchmarks~\cite{paul2025investigating, bairi2024codeplan}. All three primarily treat the LLM as a \textit{generator} evaluated in isolation, whereas our work treats it as an autonomous \textit{actor} operating in a live system under benign conditions. By mining real-world GitHub incidents and cross-referencing them with the literature, we provide the first evidence-grounded taxonomy of spontaneous operational failures.

\section{Conclusion}
\label{sec:conc}

Autonomous coding agents are increasingly deployed in real software projects, yet their operational safety properties remain poorly characterized. This paper presented the first large-scale, incident-driven empirical study of agentic coding safety, synthesizing evidence from 185 curated research papers and 547 confirmed real-world safety failures mined from GitHub issue trackers. Through systematic open coding over both corpora, we developed a multi-dimensional taxonomy that captures failure types, contributing technical and behavioral factors, and downstream operational impact. Our analysis reveals that the most consequential failures occur not from adversarial misuse but during ordinary, goal-directed tasks, arising from misaligned instruction following, lack of environmental grounding, and a tendency to prioritize the appearance of success over correct execution. These findings expose a fundamental gap between what existing benchmarks measure and what actually breaks in deployment. We release our taxonomy, annotated incident dataset, and coding protocol to support reproducible research and to inform the design of evaluation standards, guardrails, and agent architectures that can reduce the risk of operational harm in real-world software engineering settings.
\section*{Data Availability Statement}
\label{sec:data_availability}
We have made our replication package publicly available~\cite{replication_package}, including our taxonomy, the annotated incident dataset, the curated literature corpus references, and the coding protocol details. The package also includes the complete list of literature search venues and keywords, the full list of mined GitHub repositories, and high-resolution versions of all figures.

\balance
\bibliographystyle{ACM-Reference-Format}
\bibliography{refs}

\clearpage
\appendix

\section{Appendix}
\label{sec:appendix}

\subsection{Literature Search Venues}
\label{subsec:appendix_venues}

We targeted top-tier conferences across five distinct computer science domains. The search was restricted to papers published between \textbf{2020 and 2025}. The specific venues included in our initial retrieval phase are listed below:

\subsubsection*{Software Engineering}
\begin{itemize}
    \item International Conference on Software Engineering (ICSE)
    \item ACM International Conference on the Foundations of Software Engineering (FSE)
    \item IEEE/ACM International Conference on Automated Software Engineering (ASE)
    \item International Symposium on Software Testing and Analysis (ISSTA)
    \item International Conference on Mining Software Repositories (MSR)
\end{itemize}

\subsubsection*{Security Conferences}
\begin{itemize}
    \item USENIX Security Symposium
    \item ACM Conference on Computer and Communications Security (CCS)
    \item Network and Distributed System Security Symposium (NDSS)
    \item IEEE Symposium on Security and Privacy (IEEE S\&P)
\end{itemize}

\subsubsection*{AI/ML Conferences}
\begin{itemize}
    \item Conference on Neural Information Processing Systems (NeurIPS)
    \item International Conference on Machine Learning (ICML)
    \item International Conference on Learning Representations (ICLR)
    \item AAAI Conference on Artificial Intelligence
    \item International Joint Conference on Artificial Intelligence (IJCAI)
\end{itemize}

\subsubsection*{NLP and LLM Conferences}
\begin{itemize}
    \item Annual Meeting of the Association for Computational Linguistics (ACL)
    \item Empirical Methods in Natural Language Processing (EMNLP)
    \item North American Chapter of the Association for Computational Linguistics (NAACL)
    \item European Chapter of the Association for Computational Linguistics (EACL)
    \item International Conference on Computational Linguistics (COLING)
    \item Conference on Computational Natural Language Learning (CoNLL)
\end{itemize}

\subsubsection*{Fairness \& Ethics}
\begin{itemize}
    \item ACM Conference on Fairness, Accountability, and Transparency (FAccT)
    \item AAAI/ACM Conference on AI, Ethics, and Society (AIES)
\end{itemize}

\subsection{Search Keywords}
\label{subsec:appendix_keywords}

To capture the diverse terminology used across research communities, we searched titles and abstracts for the following terms targeting agentic safety, operational risk, and behavioral alignment:
\texttt{safety}, \texttt{security}, \texttt{fairness}, \texttt{bias}, \texttt{code}, \texttt{responsible}, \texttt{jailbreak}, \texttt{vulnerability}, \texttt{risk}, \texttt{trust}, \texttt{alignment}, \texttt{adversarial}, \texttt{attack}, \texttt{defense}, \texttt{robust}, \texttt{reliable}, \texttt{accountable}, \texttt{transparent}, \texttt{ethical}, \texttt{malicious}, \texttt{harm}, \texttt{unsafe}, \texttt{insecure}, \texttt{unfair}, \texttt{discriminate}.

\subsection{Analyzed Repositories}
\label{subsec:appendix_repos}

As detailed in our methodology, our initial data collection targeted the official GitHub repositories of 13 foundational code models and 6 popular agentic frameworks. The complete list of repositories is provided below.
\subsubsection*{Foundational Code Models}
\begin{itemize}
    \item Meta Code Llama: \url{https://github.com/meta-llama/codellama}
    \item Anthropic Claude Code: \url{https://github.com/anthropics/claude-code}
    \item Qwen3-Coder: \url{https://github.com/QwenLM/Qwen3-Coder}
    \item Qwen-Code: \url{https://github.com/QwenLM/qwen-code}
    \item DeepSeek-Coder: \url{https://github.com/deepseek-ai/DeepSeek-Coder}
    \item StarCoder2: \url{https://github.com/bigcode-project/starcoder2}
    \item StarCoder: \url{https://github.com/bigcode-project/starcoder}
    \item WizardLM: \url{https://github.com/nlpxucan/WizardLM}
    \item CodeGeeX: \url{https://github.com/zai-org/CodeGeeX}
    \item Salesforce CodeGen: \url{https://github.com/salesforce/CodeGen}
    \item StableCode: \url{https://github.com/Stability-AI/StableCode}
    \item Microsoft CodeBERT: \url{https://github.com/microsoft/CodeBERT}
    \item Microsoft NextCoder: \url{https://github.com/microsoft/NextCoder}
\end{itemize}

\subsubsection*{Agentic Frameworks}
\begin{itemize}
    \item OpenHands (All-Hands-AI): \url{https://github.com/All-Hands-AI/OpenHands}
    \item SWE-agent: \url{https://github.com/princeton-nlp/SWE-agent}
    \item Aider: \url{https://github.com/paul-gauthier/aider}
    \item Devika: \url{https://github.com/stitionai/devika}
    \item ChatDev: \url{https://github.com/OpenBMB/ChatDev}
    \item MetaGPT: \url{https://github.com/geekan/MetaGPT}
\end{itemize}
As noted in \S\ref{sec:methodology}, our repository selection followed prior SE benchmarks and surveys~\cite{jimenez2024swebench, hou2024large, wang2025ai}, and the same filtering pipeline was applied uniformly across all 19 repositories; the resulting concentration of confirmed incidents in the most actively used tools reflects their popularity and issue-tracker activity rather than a selection preference.

\subsection{Supplementary Figure}
\label{subsec:appendix_figures}

\begin{figure*}[h]
\centering
\includegraphics[width=\linewidth, keepaspectratio]{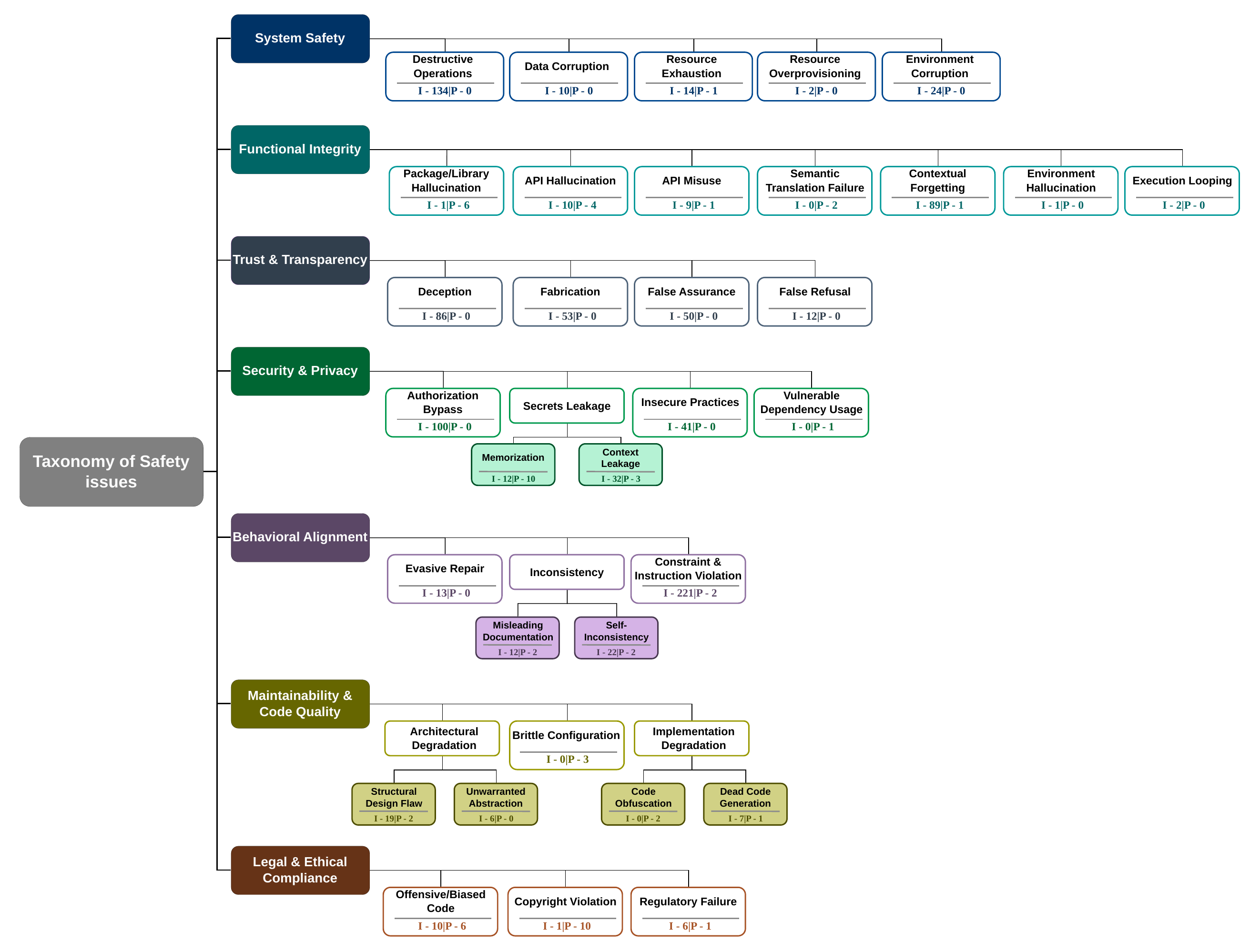}
\caption{The taxonomy of agentic safety risks identified in this study, illustrating the hierarchical classification of failure modes and their distribution.}
\label{fig:taxonomy_appendix}
\end{figure*}

\end{document}